\documentclass[letterpaper]{article}
\usepackage{aaai25}
\nocopyright
\def\myurl#1{\setbox0\vbox{\hsize.5\maxdimen}}

\AtBeginDocument{%
  \providecommand\BibTeX{{%
    \normalfont B\kern-0.5em{\scshape i\kern-0.25em b}\kern-0.8em\TeX}}}

\usepackage{times}
\usepackage{helvet}
\usepackage{courier}
\usepackage{graphicx}
\usepackage[hyphens]{url}  
\usepackage{natbib}  
\usepackage{booktabs}
\usepackage{float}
\usepackage{comment}
\usepackage{enumerate}
\usepackage{caption}
\usepackage{hyphenat}
\PassOptionsToPackage{hyphens}{url}

\usepackage{courier}
\usepackage{amsmath,amsthm, graphicx, multicol, array, mathtools}
\usepackage{color} 
\usepackage{array,multirow,graphicx}

\newcommand{\ignore}[1]{}

\usepackage{xspace}
\usepackage{amsmath}
\usepackage{amsfonts}
\usepackage{url}
\usepackage{marvosym}
\usepackage{ifthen}
\theoremstyle{plain}

\newcommand{\chatoDisplayMode}[1]{#1}

\definecolor{MyRed}{rgb}{0.6,0.0,0.0} 
\definecolor{MyBlack}{rgb}{0.1,0.1,0.1} 
\newcommand{\inred}[1]{{\color{MyRed}\sf\textbf{\textsc{#1}}}}
\newcommand{\frameit}[2]{
  \begin{center}
  {\color{MyRed}
  \framebox[.9\columnwidth][l]{
    \begin{minipage}{.85\columnwidth}
    \inred{#1}: {\sf\color{MyBlack}#2}
    \end{minipage}
  }\\
  }
  \end{center}
}

\newcommand{\note}[2][]{\chatoDisplayMode{\def\@tmpsig{#1}\frameit{{\Pointinghand} Note}{#2\ifx \@tmpsig \@empty \else \mbox{ --\em #1}\fi}}}
\newcommand{\todo}[2][]{\chatoDisplayMode{\def\@tmpsig{#1}\frameit{{\Writinghand} To-do}{#2\ifx \@tmpsig \@empty \else \mbox{ --\em #1}\fi}}}

\newcommand{\abbrevStyle}[1]{#1}

\newcommand{\ie}{\abbrevStyle{i.e.}\xspace}

\newcommand{\cf}{\abbrevStyle{cf.}\xspace}

\newcommand{\etc}{\abbrevStyle{etc.}\xspace}

\newcommand{\xhdr}[1]{\vspace{1.7mm}\noindent{{\bf #1.}}}

\newcommand{\textcite}[1]{\citeauthor{#1} \shortcite{#1}}

\newcommand{\hide}[1]{}

\makeatletter
\newcommand{\iffont}[2]{\ifthenelse{\equal{\f@family}{#1}}{#2}{}}
\makeatother

\iffont{ptm}{
  \usepackage{mathptmx}

  \DeclareSymbolFont{greek}{OML}{cmm}{m}{n}
  \DeclareMathSymbol{\alpha}{\mathalpha}{greek}{"0B}
  \DeclareMathSymbol{\beta}{\mathalpha}{greek}{"0C}
  \DeclareMathSymbol{\gamma}{\mathalpha}{greek}{"0D}
  \DeclareMathSymbol{\delta}{\mathalpha}{greek}{"0E}
  \DeclareMathSymbol{\epsilon}{\mathalpha}{greek}{"0F}
  \DeclareMathSymbol{\zeta}{\mathalpha}{greek}{"10}
  \DeclareMathSymbol{\eta}{\mathalpha}{greek}{"11}
  \DeclareMathSymbol{\theta}{\mathalpha}{greek}{"12}
  \DeclareMathSymbol{\iota}{\mathalpha}{greek}{"13}
  \DeclareMathSymbol{\kappa}{\mathalpha}{greek}{"14}
  \DeclareMathSymbol{\lambda}{\mathalpha}{greek}{"15}
  \DeclareMathSymbol{\mu}{\mathalpha}{greek}{"16}
  \DeclareMathSymbol{\nu}{\mathalpha}{greek}{"17}
  \DeclareMathSymbol{\xi}{\mathalpha}{greek}{"18}
  \DeclareMathSymbol{\pi}{\mathalpha}{greek}{"19}
  \DeclareMathSymbol{\rho}{\mathalpha}{greek}{"1A}
  \DeclareMathSymbol{\sigma}{\mathalpha}{greek}{"1B}
  \DeclareMathSymbol{\tau}{\mathalpha}{greek}{"1C}
  \DeclareMathSymbol{\upsilon}{\mathalpha}{greek}{"1D}
  \DeclareMathSymbol{\phi}{\mathalpha}{greek}{"1E}
  \DeclareMathSymbol{\chi}{\mathalpha}{greek}{"1F}
  \DeclareMathSymbol{\psi}{\mathalpha}{greek}{"20}
  \DeclareMathSymbol{\omega}{\mathalpha}{greek}{"21}
  \DeclareMathSymbol{\varepsilon}{\mathalpha}{greek}{"22}
  \DeclareMathSymbol{\vartheta}{\mathalpha}{greek}{"23}
  \DeclareMathSymbol{\varpi}{\mathalpha}{greek}{"24}
  \DeclareMathSymbol{\varrho}{\mathalpha}{greek}{"25}
  \DeclareMathSymbol{\varsigma}{\mathalpha}{greek}{"26}
  \DeclareMathSymbol{\varphi}{\mathalpha}{greek}{"27}
  \DeclareSymbolFont{otone}{OT1}{cmr}{m}{n}
  \DeclareMathSymbol{\Gamma}{\mathalpha}{otone}{0}
  \DeclareMathSymbol{\Delta}{\mathalpha}{otone}{1}
  \DeclareMathSymbol{\Theta}{\mathalpha}{otone}{2}
  \DeclareMathSymbol{\Lambda}{\mathalpha}{otone}{3}
  \DeclareMathSymbol{\Xi}{\mathalpha}{otone}{4}
  \DeclareMathSymbol{\Pi}{\mathalpha}{otone}{5}
  \DeclareMathSymbol{\Sigma}{\mathalpha}{otone}{6}
  \DeclareMathSymbol{\Upsilon}{\mathalpha}{otone}{7}
  \DeclareMathSymbol{\Phi}{\mathalpha}{otone}{8}
  \DeclareMathSymbol{\Psi}{\mathalpha}{otone}{9}
  \DeclareMathSymbol{\Omega}{\mathalpha}{otone}{10}
  \DeclareSymbolFont{syms}{OML}{cmm}{m}{it}
  \DeclareMathSymbol{\partial}{\mathord}{syms}{"40}
  \DeclareMathAlphabet{\mathbold}{OML}{cmm}{b}{it}
  \DeclareSymbolFont{largesymbols}{OMX}{cmex}{m}{n}

}

\newcommand{\ww}{\textsc{Web2Wiki}\xspace}

\usepackage{xcolor}
\newcommand{\answerYes}[1]{\textcolor{blue}{#1}} 
\newcommand{\answerNo}[1]{\textcolor{teal}{#1}} 
\newcommand{\answerNA}[1]{\textcolor{gray}{#1}}

\begin{document}
%
\title{Web2Wiki: Characterizing Wikipedia Linking Across the Web}
\author{Veniamin Veselovsky, Tiziano Piccardi, Ashton Anderson, Robert West, Akhil Arora}

\author {
    Veniamin Veselovsky\textsuperscript{\rm 1}, 
    Tiziano Piccardi\textsuperscript{\rm 2}, 
    Ashton Anderson\textsuperscript{\rm 3}, 
    Robert West\textsuperscript{\rm 4}\thanks{Robert West is a Wikimedia Foundation Research Fellow.},
    Akhil Arora\textsuperscript{\rm 5}
}
\affiliations {
    \textsuperscript{\rm 1}Princeton University \hfill \textsuperscript{\rm 2}Stanford University \hfill \textsuperscript{\rm 3}University of Toronto \hfill \textsuperscript{\rm 4}EPFL \hfill \textsuperscript{\rm 5}Aarhus University\\
    veniamin@princeton.edu, piccardi@stanford.edu, ashton@cs.toronto.edu, robert.west@epfl.ch, akhil.arora@cs.au.dk
}

\maketitle
\begin{abstract}
Wikipedia is one of the most visited websites globally, yet its role beyond its own platform remains largely unexplored. In this paper, we present the first large-scale analysis of how Wikipedia is referenced across the Web. Using a dataset from Common Crawl, we identify over 90 million Wikipedia links spanning 1.68\% of Web domains and examine their distribution, context, and function. Our analysis of English Wikipedia reveals three key findings: (1) Wikipedia is most frequently cited by news and science websites for informational purposes, while commercial websites reference it less often. (2) The majority of Wikipedia links appear within the main content rather than in boilerplate or user-generated sections, highlighting their role in structured knowledge presentation. (3) Most links (95\%) serve as explanatory references rather than as evidence or attribution, reinforcing Wikipedia’s function as a background knowledge provider. While this study focuses on English Wikipedia, our publicly released WEB2WIKI dataset includes links from multiple language editions, supporting future research on Wikipedia’s global influence on the Web.
\end{abstract}

\maketitle





\section{Introduction}

As the de facto encyclopedia of the Internet, Wikipedia is one of the most visited websites globally~\cite{why_we_read_wikipedia} and is widely cited as an authoritative source of information~\cite{mesgari2015sum}.
This unique status and prosperity have further inspired rich bodies of academic literature on Wikipedia, providing insights about its structure~\cite{holloway2007analyzing}, readership~\cite{why_the_world_reads_wikipedia}, knowledge gaps~\cite{zia2019knowledge}, as well as user behaviors~\cite{wikinav_approx,piccardi2021large,piccardi22}. 
Initial research has also examined Wikipedia's role beyond its own platform, analyzing its impact on law and science~\cite{thompson2022trial,thompson2018science}, its importance on social media~\cite{vincent2018examining}, and its influence on search engine quality~\cite{mcmahon2017substantial,vincent2019measuring}. However, the extent of Wikipedia's presence across the broader Web remains unexplored, leaving a significant gap in our understanding of its external influence.
While most of its traffic originates from search engines, a substantial portion comes via direct links to its articles~\cite{piccardi2021large}. Platforms such as YouTube and news websites integrate Wikipedia links to provide additional context for specific content or controversial topics\footnote{\url{https://support.google.com/youtube/answer/7630512}}. Similarly, users of platforms like Quora or Reddit frequently reference Wikipedia articles to support their arguments~\cite{moyer2015determining}.

Given its central role, Wikipedia should not be studied in isolation; mapping its role across the Web is critical. First, understanding how Wikipedia-generated knowledge is used beyond its own platform can reveal key insights. These include which articles are most frequently referenced, the contexts in which they appear, the types of websites linking to Wikipedia, and the motivations behind these links. 

Second, it can help quantify the extent and size of Wikipedia by measuring its reach, which is critical in assessing its value and understanding its global influence. These insights can inform policies by assessing Wikipedia's potential impact on public discourse, indicating areas where Wikipedia could play a central role in shaping public understanding.
Finally, it can open up a new paradigm for studying what matters on Wikipedia through the rich world of public links. It may help expose biases or gaps in the topics most frequently referenced, and inform improvements to better support user needs

Specifically, we formulate three research questions:
\begin{itemize}
    \item \textbf{RQ1:} \textit{What} types of Wikipedia articles are referenced most on the Web?
    \item \textbf{RQ2:} \textit{Where} are Wikipedia articles most likely to be linked?
    \item \textbf{RQ3:} \textit{Why} do people reference Wikipedia articles on external websites?
\end{itemize}

To answer these research questions, we generated a dataset called \ww containing all Wikipedia links from Common Crawl---the largest public dump of HTML content on the Web, and conducted a large-scale analysis of how English Wikipedia is linked across the Web.
We begin by examining which types of Wikipedia articles are disproportionately linked across different parts of the Web. To contextualize these patterns and provide a complete picture, we compare how frequently articles are linked on the broader Web and the social Web, with a specific focus on Reddit. Following prior work that uses in-degree as a measure of Wikipedia article importance~\cite{fortunato2006approximating,thalhammer2016pagerank}, we use the number of incoming links from Wikipedia's internal network as a baseline proxy for an article's significance.

Then, to examine where Wikipedia is linked, we conduct (1) a domain-level analysis of linking websites and (2) an in-page segmentation of Wikipedia references. Our findings support the expectation of Wikipedia as an authority for validating knowledge: Wikipedia articles are far more likely to be referenced on News, Science, and Society websites when compared to Business and Shopping websites. When Business and Shopping websites do link to Wikipedia, they are more likely to place links in the boilerplate portion of the page rather than within the main content.

Finally, to understand why Wikipedia is referenced, we identified two primary motivations through iterative coding of Wikipedia links across the Web. The predominant reason is ``evidence,'' and ``delegation'.'
Evidence-based linking occurs when a website embeds Wikipedia-sourced content verbatim or cites Wikipedia to support a claim, fact, or media resource. In contrast, delegation-based linking is broader in scope. Websites use Wikipedia for contextual explanations (e.g., linking to the Wikipedia article on HTTP cookies when explaining privacy policies), for background information on niche topics, or as an integrated reference across various content types.

Alongside these findings, we release the \ww dataset to facilitate further research on this topic, advance our understanding of Wikipedia's role on the Web, and serve as a foundation for future analyses of linking structures across the Web.

\section{Related Work}

The Wikimedia Foundation has long emphasized the importance of studying Wikipedia's broader connections to the Web. In 2015, its ``New Research Directions'' identified this as a key priority~\cite{taraborelli2015sum}.
The relative research roadmap highlights the need to determine what types of knowledge Wikimedia must acquire to better serve its role on the Web~\cite{Gaps}. This work supports that agenda by offering a descriptive analysis and releasing the first dataset that maps Wikipedia's position and usage across the Web.

\xhdr{Value to the Web}
Most prior studies have focused on specific platforms, particularly Wikipedia's role in social media and search engines.
A theme in this line of work has been quantifying the value that Wikipedia offers these sites.
Social media platforms, particularly Reddit and StackOverflow, benefit significantly from Wikipedia references, whereas Wikipedia itself receives little direct traffic in return~\cite{vincent2018examining}. This phenomenon has been described as the ``paradox of reuse'': as Wikipedia increasingly powers external technologies, users may rely on its content without ever visiting the site directly~\cite{mcmahon2017substantial}. Other studies focused on the role Wikipedia plays on search engines, finding that Wikipedia's content provides a large benefit for Google being omnipresent in search results and a critical tool for effective information retrieval due to the decrease in search result quality with Wikipedia removal~\cite{vincent2019measuring,mcmahon2017substantial}.  
Extending this research beyond Reddit, StackOverflow, and search engines, a recent dataset release contains all mentions of Wikipedia articles on Twitter, providing new opportunities for analysis~\cite{meier2022twikil}.

Additionally, Wikipedia articles being invoked in the Reddit r/TodayILearned (TIL) subreddit leads to a nontrivial increase in Wikipedia viewership \cite{moyer2015determining}, and Wikipedia is a gateway to the Web, with large swathes of Wikipedia browsing activity leading to other websites via external links present in Wikipedia articles~\cite{forte2018information,piccardi2021value}.

Finally, Wikipedia's value to the Web has been studied through the lens of plagiarism detection~\cite{alshomary2019wikipedia}. They construct a dataset of Wikipedia references extracted from Common Crawl to identify unattributed content on the Web, and then detect plagiarism by using hashing to find matching pairs of textual content between the Web and Wikipedia.

\xhdr{Societal impact beyond the Web} A limited body of research has explored Wikipedia's broader societal impact. One study conducted a randomized field experiment on legal articles in Ireland and found that judges heavily rely on Wikipedia in shaping judicial behaviour~\cite{thompson2022trial}. Another work similarly showed that Wikipedia articles are used as review pages for many scientific analyses~\cite{thompson2018science}. 
Further studies have linked Wikipedia to measurable economic effects. For example, Wikipedia content influences tourism revenue, as destinations with better Wikipedia coverage attract more visitors~\cite{hinnosaar2023wikipedia} and the value of traffic Wikipedia drives to external websites--if it were acquired through online advertising--would amount to millions of dollars~\cite{piccardi2021value}.

\section{\ww: Data and Methodology}\label{sec:data}
\xhdr{Data and code}
The dataset used in this paper was extracted from the February 2021 Common Crawl dump\footnote{\url{https://commoncrawl.org/2021/03/february-march-2021-crawl-archive-now-available/}}---the largest public scrape of HTML pages on the Web created prior to February 2021---through a regex search of {\small\ttfamily{<a>}} tags, the HTML element for hyperlinking. 
Specifically, the \ww dataset, released with this article, contains a large sample of the publicly accessible HTML pages that link to Wikimedia projects, as well as the bare webpage--article link pairs. 
Given the web-scraping policies of Common Crawl (\cf Appendix.~\ref{app:cc}), the dataset contains links to Wikipedia articles available on the surface Web---the portion of the Web accessible to the general public and searchable with standard Web search engines\footnote{https://en.wikipedia.org/wiki/Surface\_web}---, excluding websites such as Facebook, Reddit, Twitter, Quora, or YouTube that require registering an account to see their content.

The dataset will be made publicly available on Zenodo under an open license (CC-BY-SA 4.0) upon acceptance. All code required to reproduce the analyses presented in this paper has been anonymously made available at~\url{https://anonymous.4open.science/r/web2wiki-2AF8}.

\begin{table*}[t]\centering
	\small
	\begin{tabular}{lrrrrr}\toprule
        \multirow{2}{*}{\textbf{Language}} & \multirow{2}{*}{\textbf{\#Domains}}& \multirow{1}{*}{\textbf{\#Webpages with}} & \multirow{1}{*}{\textbf{\#Wikipedia articles}} & \multirow{1}{*}{\centering\textbf{\#Incoming}} & \multirow{1}{*}{\centering\textbf{\%Wikipedia articles responsible}} \\
        		& & \multirow{1}{*}{\textbf{links to Wikipedia}} & \multirow{1}{*}{\textbf{linked on the Web}} & \multirow{1}{*}{\textbf{Web-links}} & \multirow{1}{*}{\textbf{for 80\% of the Web-links}}\\
  \midrule
		English (EN)      & 940,239         & 14,462,267        & 3,619,416                           & 48,829,702              & 9\%          \\
		German (DE)       & 155,887         & 2,128,814         & 824,981                             & 5,882,838               & 17\%          \\
		French (FR)       & 96,875          & 1,465,997         & 692,220                             & 4,925,889               & 17\%          \\
		Japanese (JA)     & 56,651          & 818,110        & 656,086                             & 4,455,828               & 20\%          \\
		Spanish (ES)      & 88,903          & 1,262,583        & 486,811                             & 3,629,491               & 17\%          \\
		Russian (RU)      & 53,219          & 1,105,698         & 465,283                             & 2,705,021               & 19\%          \\
		Italian (IT)      & 55,329          & 851,533         & 350,046                             & 2,463,232               & 16\%          \\
		Portuguese (PT)   & 27,350          & 413,034         & 226,713                             & 1,194,040               & 22\%          \\
		Swedish (SV)      & 14,716          & 256,776           & 221,530                             & 1,138,989               & 23\%          \\
		Dutch (NL)        & 33,953          & 366,933           & 206,727                             & 892,750                 & 29\%          \\
		Vietnamese (VI)   & 11,190          & 146,623           & 101,427                             & 882,410                 & 12\%          \\
		Polish (PL)       & 23,012          & 376,167          & 161,395                             & 752,888                 & 23\%          \\
		\bottomrule
	\end{tabular}
	\caption{Summary statistics of the \ww dataset, portraying the number of distinct domains and webpages that link to Wikipedia, the number of distinct Wikipedia articles that are linked on the Web, the total number of incoming Web-links to Wikipedia, and the percentage of Wikipedia articles that are responsible for 80\% of the Web-links, respectively, across the 12 most linked language versions of Wikipedia on the Web.}
	\label{tab:data_description}
\end{table*}

\subsection{Data pre-processing and summary}
To conduct the analyses presented in this paper, the first step involves processing each webpage to retrieve all Wikipedia links found within an {\small\ttfamily{<a>}} tag. Note that by restricting our focus solely to links that point to Wikipedia in English, we exclude the links to all other Wikimedia projects (\textit{Wikiquote, Wikinews, Wikiversity, Wiktionary, Wikisource, Wikibooks, Wikivoyage}) from our analyses. However, these links are included in the full \ww dataset. 
Starting from a dataset of 2.7 billion web pages—amounting to 280 TiB of uncompressed content—our pre-processed dataset contains 90,805,367 links to Wikipedia articles, spanning 1.68\% of the domains in the Common Crawl dump. 

Diving deeper into the dataset statistics (Table~\ref{tab:data_description}), English-language Wikipedia stands out for several reasons. First, it dominates the \ww dataset, accounting for over 50\% of all links, meaning it is referenced almost 10 times more frequently than the next most-used language (German).
Second, links to English Wikipedia are highly concentrated: 80\% of incoming links target just 9\% of English articles---a significantly more skewed distribution than observed in other languages.
Third, a large proportion (58\%) of English Wikipedia articles are linked on the broader Web, compared to German (33\%), French (30\%), Spanish (30\%), and Dutch (10\%). Given these factors--along with the authors' proficiency in English--we focus our analysis on English Wikipedia, which corresponds to 48,829,702 links from 14,462,267 distinct webpages across 940,239 domains.

Overall, the distribution of Wikipedia links per webpage follows a long-tail pattern, with an average of 3.4 links per page, while some pages contain over 4,000 Wikipedia links.

\subsection{Methodology} We focus on three important questions (\emph{what}, \emph{where}, and \emph{why}) to ground our understanding of the true extent of Wikipedia's presence on the Web.

\xhdr{Measuring article importance}
We first address which Wikipedia articles are most salient on the Web by comparing their presence in general web links to their use on the social Web, specifically in Reddit posts--a platform not included in Common Crawl data. Our data consists of all comments and submissions on Reddit from its inception in December  2005  to  June  2022 \cite{baumgartner2020pushshift}. This includes almost  13  billion comments by  89  million users and 5 billion submissions by 48 million users. 
To contextualize these patterns, we use Wikipedia's in-degree as a baseline proxy for article importance within the platform. We compute in-degrees from the Wikipedia XML dump released concurrently with the Common Crawl data, considering only internal links present in the body of each article~\cite{mitrevski2020wikihist}. This measure serves as a useful reference point, as it implicitly reflects an article's significance in Wikipedia's internal link network~\cite{fortunato2006approximating,thalhammer2016pagerank}.

We then analyze which ``types'' of Wikipedia articles are disproportionately referenced on both the broader Web and Reddit.

\xhdr{ORES topics}
To obtain a broader classification of Wikipedia articles that are highly represented on the Web, we use Wikipedia's ORES topic model~\cite{halfaker2020ores}.
ORES classifies articles by topic and clusters them into sixty-four categories\footnote{\url{https://www.mediawiki.org/wiki/ORES/Articletopic}}, grouping articles into broad themes such as \textit{computing}, \textit{biology}, and \textit{society}.

\xhdr{Websites topics}
Next, we examine where Wikipedia articles are linked by analyzing the topics of websites that reference Wikipedia.
To classify webpage topics, we use Homepage2Vec~\cite{lugeon2022homepage2vec}, a model that predicts a webpage's topic based on its content. This classification follows Curlie's taxonomy\footnote{\url{https://curlie.org/}}, which defines 14 distinct topics, including \textit{News}, \textit{Science}, and \textit{Shopping}.

\xhdr{Webpage segmentation}
We then analyze where \textit{within} a webpage, the Wikipedia article link appears. To achieve this, we manually define a set of structural rules to segment webpages into three distinct sections: boilerplate, main content, and user contributions. To develop these segmentation rules, the authors manually inspected 500 random webpages containing Wikipedia links.

Boilerplate includes static elements that remain consistent across different pages, such as headers and footers. The main content encompasses the primary text of a webpage, where most of the substantive information is presented. Finally, user contributions capture Wikipedia links shared through interactive user-generated content, such as those appearing in comment sections and discussion threads.

\section{What Articles Are Linked}\label{what}

The first question we examine is what types of Wikipedia articles are linked across the Web. A straightforward approach would be to count the number of distinct domains linking to a given article. However, interpreting these counts requires a meaningful baseline. As introduced before, we use Wikipedia in degrees to contextualize web links as a reference measure. This baseline assumes that an article is important if it is heavily referenced within Wikipedia itself. By comparing Wikipedia links on the general Web and Reddit to in-links within Wikipedia, we quantify how external references differ from Wikipedia's internal linking structure--comparing semantic connections within Wikipedia with how articles are referenced externally on the broader Web and the social Web (Reddit).

\begin{figure}[!t]
    \centering
    \includegraphics[width = 3.3in]{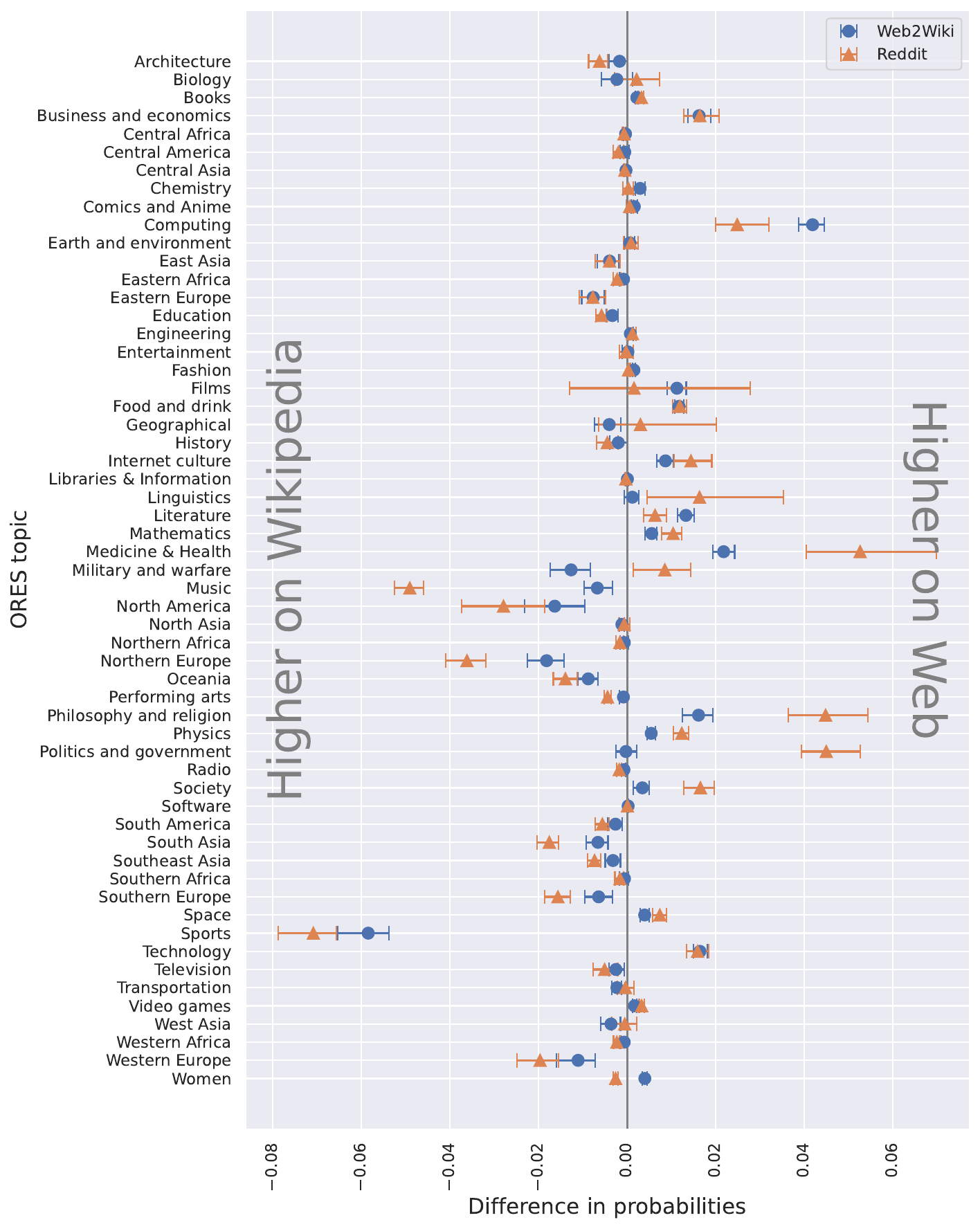}
    \caption{Comparison of proportion of Wikipedia-internal links, Web links, and Reddit shares of Wikipedia. A higher score means proportionally more in-links from external websites than as Wikipedia-internal links. Error bars represent 95\% bootstrapped confidence intervals.}
    \label{fig:ores_topics}
\end{figure}

The Pearson correlation between external website links and Wikipedia article in-degree (number of incoming links) shows a significant positive relationship ($r = 0.5,~p<0.001$), indicating that articles with high in-degree (\ie, those referred by many other concepts) are more likely to be linked from external websites.
By comparing how the three groups--the Web, Reddit, and Wikipedia itself--link to distinct topics, we identify both alignments and divergences in their referencing patterns.

Wikipedia-internal links, external web references, and Reddit citations each have distinct distributions over ORES topics. To quantify these distributions, we obtained the probability distribution--over the 64 topics--of Wikipedia articles linked from each source.

We define the \textit{Wikipedia-internal probability} of a topic $k$ as the likelihood that a randomly selected Wikipedia internal link points to an article on that topic. Similarly, we define the \textit{Web probability} as the probability that an external website links to a Wikipedia article in a given ORES topic, and the \textit{Reddit probability} as the probability that a Wikipedia article cited on Reddit belongs to a specific ORES topic.
By comparing these three probability distributions, we identify which topics are disproportionately referenced on the Web and Reddit relative to Wikipedia's internal linking structure. This analysis reveals differences in how various platforms emphasize certain topics, shedding light on external vs. internal knowledge representation.

Figure \ref{fig:ores_topics} illustrates the differences in topic distributions across Wikipedia-internal links, Web links, and Reddit shares. The $x$-axis represents the probability difference for a given ORES topic's Wikipedia-internal probability and either its Web probability (blue points) or Reddit probability (orange triangles).
We observe that certain topics--such as Computing, Medicine \& Health, Philosophy \& Religion, and Technology--are referenced more frequently on the Web than Wikipedia's internal linking. In contrast, topics like Sports, North America, Northern Europe, Military, and Music appear underrepresented in external links relative to their presence in Wikipedia's internal network. While some topic-level differences in link probabilities may appear numerically small, the scale of our dataset means that even modest differences translate into substantial differences in exposure and referencing across the Web.

\xhdr{Key Findings} Wikipedia article referencing patterns vary across platforms: articles that receive frequent internal links within Wikipedia differ from those that are widely linked on the Web or shared on Reddit. This indicates that social media is not a reliable proxy for broader Wikipedia usage. While previous studies have focused on Reddit and Twitter, our findings suggest that social media linking behavior does not accurately represent how Wikipedia is referenced across the wider Web.

\section{Where Articles Are Linked}\label{where}

First, we analyze which types of websites are more likely to link to Wikipedia. To do this, we compare webpages that contain Wikipedia links with a random sample of webpages from the general Web. This allows us to determine whether Wikipedia is disproportionately referenced in certain types of websites.

Second, we examine where within a webpage Wikipedia links appear. We categorize webpage content into three sections--boilerplate, main content, and user contributions--and define HTML-based rules to extract these segments computationally. This helps identify the context and intent behind Wikipedia links.

Finally, we combine these two analyses to explore how different websites use Wikipedia links across different webpage sections. This integrated approach provides a clearer picture of Wikipedia's role in online information dissemination.

\subsection{Wikipedia vs. the Web} 
With over 48 million links to English Wikipedia across 940,000 distinct domains, Wikipedia has a strong presence on the Web. Notably, 66\% of these domains contain Wikipedia links on multiple webpages, while 3.2\% of domains include Wikipedia links on over 100 different webpages.
At the individual webpage level, we observe that 3.8 million webpages contain multiple Wikipedia links.

Beyond these aggregate statistics, it is important to characterize the types of websites that link to Wikipedia and determine whether certain website categories disproportionately reference Wikipedia.
To understand which types of websites link to Wikipedia, we apply simple heuristic rules.  Specifically, we define a Wikipedia-related page as any webpage containing ``wiki'' or ``pedia'' in its URL, and a blog as any webpage with a date in its URL. Using these rules, we find that 11\% of Wikipedia links (4.5 million) originate from encyclopedias or Wiki-style pages, while blogs account for 29\% (11.8 million) of all Wikipedia links.

These heuristics provide a basic categorization but do not fully capture the diversity of websites linking to Wikipedia. To improve classification, we use Homepage2Vec, a model that predicts a webpage's topic based on its HTML content. By comparing Homepage2Vec classifications for Wikipedia-linking websites against a random sample of general websites, we quantify which types of websites are most likely to reference Wikipedia.

\begin{figure}[t!]
    \centering
    \includegraphics[width = 3.3in]{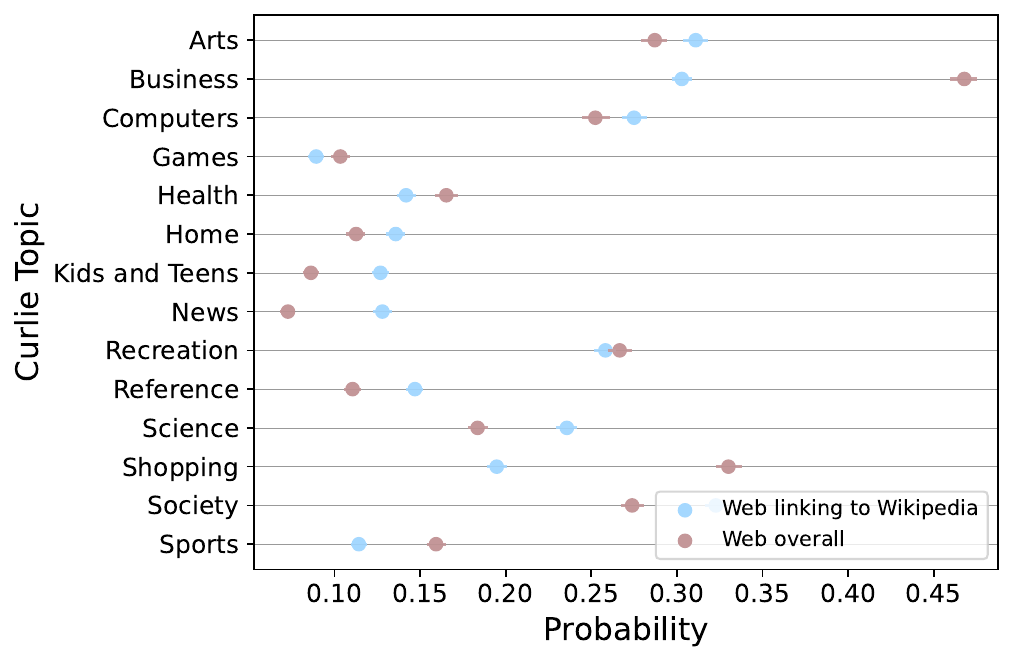}
    \caption{Comparison between websites that link to Wikipedia and a random sample of all websites. For each topic, the difference between the ``Web overall'' and ``Web linking to Wikipedia'' probabilities determines its leaning. When the score for ``Web linking to Wikipedia'' is higher, it implies that the topic tends to link more to Wikipedia, while the reverse implies that Wikipedia articles are underrepresented across that particular topic. Error bars represent 95\% bootstrapped confidence intervals.}
    \label{fig:z_score_curlie}
\end{figure}

\xhdr{Implementation and analysis}
We apply Homepage2Vec to a sample of 10,000 random domains that link to Wikipedia and 10,000 randomly selected Common Crawl domains. The model assigns a probability distribution across 14 topic categories. Since each webpage can belong to multiple categories, these probabilities do not necessarily sum to 1.

Figure \ref{fig:z_score_curlie} compares the average topic distributions of Wikipedia-linking websites and a random sample of the Web. Our analysis reveals distinct differences between websites that link to Wikipedia and the broader Web. Business and Shopping websites make up a larger portion of the Web but are less likely to link to Wikipedia articles. In contrast, News, Science, Society, Arts, Computers, and Reference websites show a higher tendency to include Wikipedia links.

A qualitative examination of these topics reveals different linking behaviors. News and Science pages often include Wikipedia links within their main content or in user-generated sections--such as on-site comments. In contrast, Business and Shopping websites are more likely to reference Wikipedia in boilerplate content, either linking to their own Wikipedia page or explaining concepts like cookies.
These findings motivate the next section of our study, where we analyze how Wikipedia links are embedded within webpage HTML structures.

\begin{table*}[!t]
    \centering
    \begin{tabular}{lrrrrr}
        \toprule 
        \textbf{Segment} &  \textbf{Webpages (M)}&  \textbf{\% of links} & \textbf{Article Entropy} & \textbf{Language Entropy} \\
        \hline 
        Boilerplate & 3.4 & 7\% & 7.6 & 7.8\\
        Main content & 43.7 &91\% & 8.4 & 8.1\\
        User-Generated & 1.0 & 2\% & 8.5 &7.5 \\%
        \hline
    \end{tabular}
    \caption{Summary of the three segments showing the number of webpages, percentage of all links, article, and language entropy.}
    \label{tab:order_estimates}
\end{table*}

\subsection{Webpage Segmentation}

Webpages are structured into different segments, each serving a distinct function that influences how links are placed and interpreted~\cite{kohlschutter2010boilerplate}. Segmenting a webpage is critical for understanding the intent behind a link's placement, as different sections are populated with links for specific reasons. For instance, links in boilerplate content typically provide site-wide navigation or general information, whereas links in main content are often used to enrich specific textual explanations. User-generated content, such as comments, includes links added by users in discussions, often for fact-checking or providing additional context.

To systematically analyze link placement, we developed a set of structural rules to identify different webpage segments. Using an iterative approach inspired by grounded theory, we continuously sampled new examples and refined the rules through deliberation among four of the authors, as detailed in Appendix~\ref{app:rules}. Applying these rules to our dataset, we find that Wikipedia links are primarily embedded in main content (91\%), while boilerplate content accounts for 7\% and user-generated content for 2\% (Table~\ref{tab:order_estimates}). This distribution highlights that Wikipedia is most frequently cited as an informational source within core webpage content, rather than in navigation menus or user discussions.

To further understand how different segments reference Wikipedia, we compute the Shannon entropy of Wikipedia articles linked in each segment. Entropy measures diversity, where higher entropy indicates a wider variety of linked articles, while lower entropy suggests a narrower, more predictable set of links.
For each segment, we randomly sample 10,000 Wikipedia articles linked from that segment and compute their probability distribution. These entropy values, reported in Table~\ref{tab:order_estimates}, show that boilerplate content has the lowest entropy, suggesting that a small, recurring set of Wikipedia articles is frequently referenced, such as explanations of standard terms like ``cookies'' or ``privacy policy.'' In contrast, main content and user-generated content have higher entropy, meaning that a more diverse range of Wikipedia articles is linked in these sections.

Since entropy calculations scale with the number of variables, we limit our analysis to 10,000 articles per segment to maintain computational efficiency. Additionally, we focus on relative entropy values rather than absolute ones to effectively compare diversity across segments.

\xhdr{Linguistic comparison}
To examine differences in webpage content, we conduct a linguistic analysis of the text surrounding Wikipedia links in each segment. For this analysis, we extract a 150-character context window on both sides of the link while masking the anchor text. We then filter out stopwords, retain the 5,000 most frequent words, and apply a TF-IDF vectorizer to compare word distributions across the three segments.

Table~\ref{tab:textual_features} presents the top 10 most predictive words for each segment, identified using a logistic regression model (\textit{one-vs-all}) on TF-IDF representations. The results show notable differences in language use. Boilerplate content is more focused on specific concepts than the other two segments, while user-generated content (e.g., comments) tends to be more conversational, featuring words such as ``look'', ``actually'', and ``here''.

To further analyze variation in textual contexts, we compute the entropy of the language used in each segment. Entropy values, reported in Table~\ref{tab:order_estimates}, reveal that main content has the highest entropy (8.1), followed by boilerplate (7.8) and user-generated content (7.5). This indicates that main content and user-generated content have similar levels of lexical diversity, whereas boilerplate content is more restricted in the variety of words used.

The lower entropy in user-generated content suggests that when Wikipedia links appear in responses or comments, they tend to be referenced in more predictable ways, with limited variation in phrasing. Similarly, the lower entropy in boilerplate content compared to main content indicates that Wikipedia links in boilerplate sections are used in consistent and repetitive contexts, such as defining terms or linking to standard references.

\begin{table}[!t]
    \centering
    \begin{tabular}{lll}
        \toprule
         \textbf{Boilerplate} & \textbf{Main content} & \textbf{User-Generated}  \\
         \toprule 
         privacy & share & reply \\
         powered & new & look \\
         twitter & second & actually \\
         yelp & time & thanks \\ 
         archive & social & think \\ 
         acknowledge & file & check \\ 
         copyright & told & link \\ 
         buy & took & example \\ 
         exhaustion & available & good \\
         designed & building & interesting \\
         \toprule
    \end{tabular}
    \caption{The 10 most predictive words characterizing each of the webpage segments.}
    \label{tab:textual_features}
\end{table}

\subsection{Topic Variations in Wikipedia Link Placement} 
The topic of a website is associated not only with its likelihood of linking to Wikipedia but also with where Wikipedia links appear within a webpage. To investigate this, we analyze how different website categories distribute Wikipedia links across boilerplate, main content, and user-generated content (e.g., comments).

This analysis builds on previous observations that boilerplate links are more common in business websites, while user-generated links are more prevalent in forums. To quantify these patterns, we map the Homepage2Vec probability distribution over Curlie topics onto webpage segments, characterizing how different website categories place Wikipedia links. Specifically, for each webpage segment, we compute the mean probability distribution of its predicted Curlie topics (from Homepage2Vec).

\begin{figure}[!t]
    \centering
    \includegraphics[width = 3.3in]{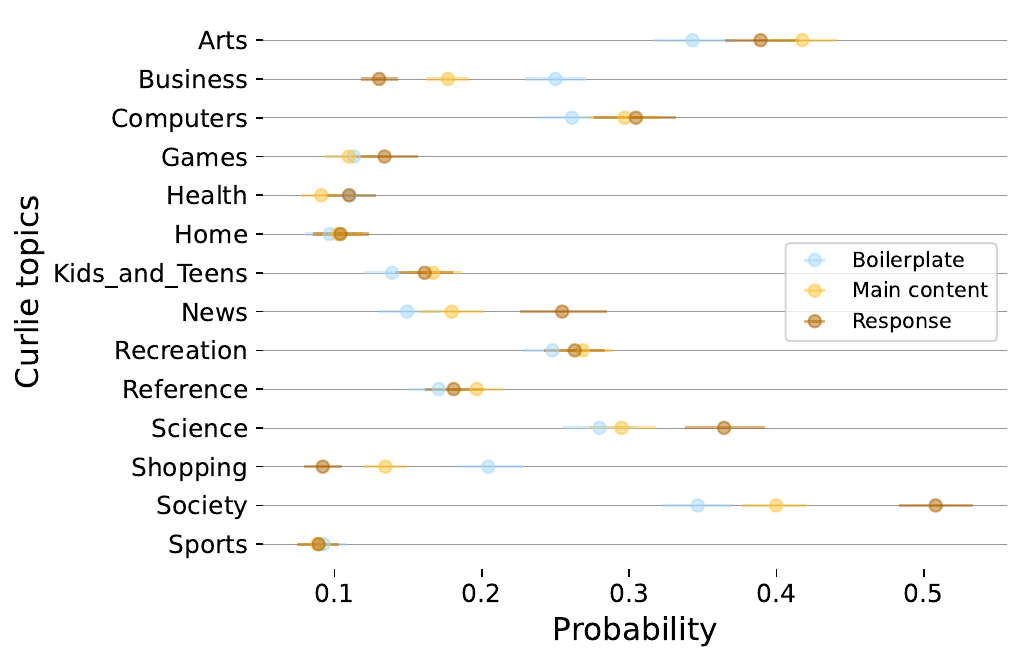}
    \caption{Comparison of where Wikipedia links are likely to occur in a webpage and their Curlie topic. Recall that Homepage2Vec outputs a separate prediction for each Curlie topic, so the probabilities across topics do not necessarily add to 1. Error bars represent 95\% bootstrapped confidence intervals.}
    \label{fig:where_and_where}
\end{figure}

Figure~\ref{fig:where_and_where} illustrates these distributions, revealing notable differences across website categories. The most significant variations occur in Arts, Business, Science, Shopping, and Society websites, suggesting distinct linking behaviors based on the site's purpose.

Among all website categories, Business and Shopping sites are the most likely to place Wikipedia links in boilerplate content. This suggests that while these websites may not reference Wikipedia frequently in their main content, they often use it in boilerplate sections, such as to explain technical terms like ``cookies.'' In contrast, Science and News websites are more likely to place Wikipedia links in main content and user-generated sections, indicating a different usage pattern where Wikipedia serves as a source of additional context or citation within discussions and articles.

These findings highlight that Wikipedia links serve multiple roles across the Web. The key distinction between websites that embed Wikipedia links in boilerplate versus main content or user-generated content appears to be driven by how business-oriented the site is. Business and commercial websites often use Wikipedia for functional or explanatory purposes, whereas informational websites such as Science and News domains rely on Wikipedia more as a content reference.

\xhdr{Key Findings}
Wikipedia is disproportionately linked from News, Science, and Society websites, while it is underrepresented in Business and Shopping websites compared to the Web as a whole. When Business and Shopping websites do link to Wikipedia, the references are primarily placed in boilerplate content, often serving functional purposes such as explaining terms like ``cookies.'' In contrast, Science and News websites are more likely to embed Wikipedia links within main content and user-generated content, suggesting that Wikipedia plays multiple roles depending on the website's context. Despite these variations, Wikipedia is most frequently linked from the main content of webpages overall.

\section{Why Articles Are Linked}\label{why}

Thus far, we have examined what Wikipedia articles are linked and where these links appear. While this provides insights into Wikipedia's presence on the Web, it does not address the intent behind including Wikipedia links. This section explores \textit{why} Wikipedia articles are referenced online.

To determine intent, we conducted iterative coding to classify the motivations behind linking to Wikipedia (\cf Appendix~\ref{app:taxonomy} for details). While there may be a wide range of reasons, our analysis reveals that most links fall into two broad categories: delegation and evidence.

\begin{itemize}
    \item \emph{Delegation} occurs when a website links to Wikipedia in a manner similar to Wikipedia's internal linking guidelines\footnote{\url{https://en.wikipedia.org/wiki/Wikipedia:Manual_of_Style/Linking#General_points_on_linking_style}}. These links serve as contextual references, directing users to relevant Wikipedia articles for additional information. Examples include linking to the Wikipedia page for HTTP Cookies when explaining a website's privacy policies or referencing an obscure place or historical figure mentioned in passing. Such links enrich content by providing supplementary explanations, defining technical terms, or offering background information.
    \item \emph{Evidence} refers to cases where content is sourced directly from Wikipedia. This category includes multimedia elements such as images, audio clips, or videos, as well as text-based references where Wikipedia is cited as the source of a fact, statistic, or quotation. Wikipedia images are commonly linked in biographical or geographic webpages, while text-based references frequently appear in blogs or informational articles where authors cite Wikipedia to support a claim.
\end{itemize}

These two categories capture the dominant linking behaviors observed across the Web: using Wikipedia as an external reference to delegate explanations or as a source to provide attribution and supporting evidence. Manually annotating samples, we were able to bucket almost all the Wikipedia sharing links (98\%) into these two motivations.

To quantify the prevalence of each type of Wikipedia linking, we randomly sampled 500 webpages that link to Wikipedia and manually annotated each link as either delegation or evidence.
Our analysis revealed that delegation is the dominant reason for linking to Wikipedia, accounting for 95\% of links, while only 5\% are used for evidence. Generalizing these proportions to the 48.8 million Wikipedia links in the main content, we estimate that 46.4 million links serve as delegation, while 2.4 million links are used for evidence.

\xhdr{Key Results}
The vast majority of Wikipedia links (95\%) are used for delegation, where websites reference Wikipedia to explain a concept. However, evidence citation accounts for 5\% of links, representing a small but meaningful portion of Wikipedia's presence on the Web.

\section{Discussion}
\label{sec:discussion}

This paper presents the first large-scale dataset of Wikipedia links across the Web, capturing the raw HTML content of webpages that reference Wikipedia articles. The dataset spans a significant portion of the Web, covering links from 1.68\% of all domains in the Common Crawl dump. By making this dataset publicly available, we provide a valuable resource for studying how Wikipedia is integrated into the broader information ecosystem.

\subsection{Summary of Findings}
Our analysis reveals several key insights into how Wikipedia is linked across the Web.
First, Wikipedia linking patterns align more closely with an article's in-degree importance within Wikipedia than with social media-driven behaviors. This distinction highlights the nuanced role Wikipedia plays online, where its articles are referenced primarily for their significance rather than for trends emerging on platforms like Reddit.

Second, the types of Wikipedia articles linked on the Web, cited on social media, and structurally significant within Wikipedia differ significantly.  The broader Web tends to link to Computing, Medicine \& Health, and Technology, while social media platforms such as Reddit are more likely to reference Philosophy and Politics. In contrast, Wikipedia's own internal link structure places greater emphasis on Sports, Music, North America, and Northern Europe, reflecting differences in external usage vs. internal network structure.

Third, Wikipedia's presence is not uniform across the Web, and different sections of webpages integrate Wikipedia links in distinct ways. Boilerplate content often includes Wikipedia links to define or explain concepts essential to the website, whereas main content links are more common in News websites, where Wikipedia is used to provide contextual references.

Finally, we introduce a systematic approach to studying linking behavior on the Web, classifying links into two fundamental categories: evidence and delegation. By analyzing a random sample of webpages, we estimate the prevalence of each linking type. Importantly, this framework is broadly applicable, allowing for the study of linking patterns across different types of websites, languages, and online ecosystems.

\subsection{Implications and Broader Impact}
Our findings reinforce Wikipedia's critical role beyond direct page views, highlighting its function as a form of knowledge infrastructure---serving both as a content source and as a mechanism for delegating explanations across millions of webpages. This underscores the need for the Wikipedia community to maintain and improve the platform, as its influence extends far beyond its own ecosystem. Wikipedia helps shape the dissemination of factual information across the Web, particularly in domains such as news and science, where it is cited most frequently

A central theme of this work is quantifying Wikipedia's value to the broader Web. By better understanding its reach and influence, we can inform strategies for its development to account for diverse user needs and maximize its societal impact.

The release of this dataset opens multiple avenues for future research. It can be used to study the credibility of sources linking to Wikipedia, examine Wikipedia's role in misinformation correction, or track changes in linking behavior over time by integrating additional datasets. Furthermore, it enables investigations into how Wikipedia contributes to knowledge dissemination across different domains and how its presence on the Web evolves in response to external events.

Beyond analyzing Wikipedia's role, this work has major implications for improving Wikipedia itself. Most studies of Wikipedia have focused on isolated analyses of its internal structure, but the \ww dataset expands the scope of research by offering a broader, web-scale perspective. Additionally, the methodology presented in this study can be adapted to analyze linking patterns beyond Wikipedia, providing a framework for studying how other entities and websites integrate into the Web's information ecosystem.

\subsection{Limitations and Future Work}

Our study is subject to several limitations. Web data, including Common Crawl, is inherently noisy, making webpage segmentation a challenging task with no universally established approach, even for extracting elements like boilerplate content. We rely on heuristics to segment webpages, which, while not perfect, have been validated through reliability checks to ensure robustness.

Due to practical constraints, we use a single Common Crawl snapshot, meaning our findings may be influenced by data variability across different snapshots. Future work should examine robustness across multiple snapshots to assess consistency over time. Additionally, while Common Crawl is the largest publicly available collection of surface Web data, it excludes platforms that require login access, such as Twitter, Reddit, or that enforce no-follow policies. As a result, our \ww dataset does not capture Wikipedia links from these websites.

Several promising directions for future research remain. The language surrounding Wikipedia links often reveals unexpected conceptual connections. For instance, links to \textit{Raphael's School of Athens} frequently appear in discussions about the ancient Academy, suggesting lateral relationships between entities~\cite{locei}. Similarly, hypertext structures introduce unique human associations, such as the Wikipedia article Proposed referendum on United Kingdom membership of the European Union featuring the phrase ``11,757-word behemoth'' as hypertext. These contextual relationships could inform link prediction models, RAG training, entity candidate generators~\cite{pti}, and entity linking models~\cite{eigenthemes, nelite}, similar to prior work \cite{spitkovsky2012cross,singh2012wikilinks}.

Another promising direction is to investigate the alignment between the number of incoming links, user clickstreams, the importance of these websites, and the actual attention received by linked articles. This would help identify potential mismatches between how often an article is referenced and how often it is accessed, extending prior work on the misalignment between demand and supply of attention on Wikipedia~\cite{warncke2015misalignment}.

Finally, a key direction for future research is to expand the analysis beyond English Wikipedia. While this study focused on English Wikipedia as a starting point, our dataset includes many other language editions. Exploring multilingual Wikipedia linking behavior can reveal cultural differences in citation practices. A preliminary analysis suggests that some languages exhibit far higher view-to-Web-link ratios than others, indicating possible differences in how Wikipedia is perceived across linguistic and cultural contexts. Investigating these patterns could offer valuable insights into global perspectives on Wikipedia's role as a knowledge source.

\subsection{Ethical Considerations}
\label{sec:ethics}

While the \ww dataset is constructed entirely using publicly available resources, the Web is an untamed resource and likely contains personally identifiable information and foul or offensive content. To alleviate this, our data release only consists of the linking structure and the surrounding context as opposed to the entire webpages. Moreover, our dataset does not contain any private or sensitive information about the browsing behavior of readers on the Web. Finally, no author was paid by the Wikimedia Foundation to conduct the analyses presented in this paper, thereby mitigating the potential for conflicts of interest in the analyses.

\section{Conclusion}
In this work, we introduced the \ww dataset and conducted the first large-scale analysis of how Wikipedia is linked across the Web, providing foundational estimates of its presence and influence on the surface Web. Our findings show that 1.68\% of all domains (1,475,057) on the surface Web link to Wikipedia, underscoring its global reach and significance.

Wikipedia plays a critical role in the Web's infrastructure, serving as both a de facto external reference for explanations and a frequent source for content and evidence. This highlights its dual function as a knowledge hub and a content provider across diverse online platforms.

The datasets, methods, and insights presented in this study contribute to a deeper understanding of Wikipedia's role on the Web. We hope this work provides a foundation for future research into Wikipedia's impact, as well as broader investigations into web-scale linking structures beyond Wikipedia.

\section*{Acknowledgements}
We thank Leila Zia, Isaac Johnson, Martin Gerlach, Manoel Horta Ribeiro, and Marko Čuljak for insightful discussions and valuable inputs. West’s lab is partly supported by grants from the Swiss National Science Foundation (200021\_185043 and 211379), Swiss Data Science Center (P2208), H2020 (952215), and Microsoft  Swiss  Joint  Research  Center. Arora’s lab is partly supported by grants from the Novo Nordisk Foundation (NNF24OC0099109), the Pioneer Centre for AI, and EU Horizon 2020 (101168951). We also gratefully acknowledge generous gifts from Facebook, Google, and Microsoft.

\bibliography{sources}

\subsection*{ICWSM Paper Checklist}

\begin{enumerate}
\item For most authors...
\begin{enumerate}
    \item  Would answering this research question advance science without violating social contracts, such as violating privacy norms, perpetuating unfair profiling, exacerbating the socio-economic divide, or implying disrespect to societies or cultures?
    \answerYes{Yes.}
  \item Do your main claims in the abstract and introduction accurately reflect the paper's contributions and scope?
    \answerYes{Yes.}
   \item Do you clarify how the proposed methodological approach is appropriate for the claims made? 
    \answerYes{Yes. Section~\ref{sec:data} introduces the methods, whereas Sections~\ref{what},~\ref{where}, and~\ref{why} further provides analysis-specific details.}
   \item Do you clarify what are possible artifacts in the data used, given population-specific distributions?
    \answerYes{Yes. Please see Data and Methodology (Section~\ref{sec:data}).}
  \item Did you describe the limitations of your work?
    \answerYes{Yes. Please see Section~\ref{sec:discussion}.}
  \item Did you discuss any potential negative societal impacts of your work?
    \answerYes{Yes. Please see Section~\ref{sec:ethics}. Overall, we do not foresee any significant negative impact.}
      \item Did you discuss any potential misuse of your work?
    \answerYes{Yes. Please see Section~\ref{sec:ethics}. Overall, we do not foresee any significant negative impact.}
    \item Did you describe steps taken to prevent or mitigate potential negative outcomes of the research, such as data and model documentation, data anonymization, responsible release, access control, and the reproducibility of findings?
    \answerYes{Yes. Please see Data and Methodology (Section~\ref{sec:data}) and Discussion (Section~\ref{sec:discussion}).}
  \item Have you read the ethics review guidelines and ensured that your paper conforms to them?
    \answerYes{Yes.}
\end{enumerate}

\item Additionally, if your study involves hypotheses testing...
\begin{enumerate}
  \item Did you clearly state the assumptions underlying all theoretical results?
    \answerNA{N/A}
  \item Have you provided justifications for all theoretical results?
    \answerNA{N/A}
  \item Did you discuss competing hypotheses or theories that might challenge or complement your theoretical results?
    \answerNA{N/A}
  \item Have you considered alternative mechanisms or explanations that might account for the same outcomes observed in your study?
    \answerNA{N/A}
  \item Did you address potential biases or limitations in your theoretical framework?
    \answerNA{N/A}
  \item Have you related your theoretical results to the existing literature in social science?
    \answerNA{N/A}
  \item Did you discuss the implications of your theoretical results for policy, practice, or further research in the social science domain?
    \answerNA{N/A}
\end{enumerate}

\item Additionally, if you are including theoretical proofs...
\begin{enumerate}
  \item Did you state the full set of assumptions of all theoretical results?
    \answerNA{N/A}
	\item Did you include complete proofs of all theoretical results?
    \answerNA{N/A}
\end{enumerate}

\item Additionally, if you ran machine learning experiments...
\begin{enumerate}
  \item Did you include the code, data, and instructions needed to reproduce the main experimental results (either in the supplemental material or as a URL)?
    \answerNA{N/A}
  \item Did you specify all the training details (e.g., data splits, hyperparameters, how they were chosen)?
    \answerNA{N/A}
     \item Did you report error bars (e.g., with respect to the random seed after running experiments multiple times)?
    \answerNA{N/A}
	\item Did you include the total amount of compute and the type of resources used (e.g., type of GPUs, internal cluster, or cloud provider)?
    \answerNA{N/A}
     \item Do you justify how the proposed evaluation is sufficient and appropriate to the claims made? 
    \answerNA{N/A}
     \item Do you discuss what is ``the cost`` of misclassification and fault (in)tolerance?
    \answerNA{N/A}
\end{enumerate}

\item Additionally, if you are using existing assets (e.g., code, data, models) or curating/releasing new assets, \textbf{without compromising anonymity}...
\begin{enumerate}
  \item If your work uses existing assets, did you cite the creators?
    \answerYes{Yes}
  \item Did you mention the license of the assets?
    \answerYes{Yes}
  \item Did you include any new assets in the supplemental material or as a URL?
    \answerNo{No. The collected \ww dataset is quite large to be shared with the submission, and sharing via a URL risks breaching the anonymity. Thus, we will release the dataset publicly via Zenodo upon acceptance.}
  \item Did you discuss whether and how consent was obtained from people whose data you're using/curating?
    \answerNA{N/A}
  \item Did you discuss whether the data you are using/curating contains personally identifiable information or offensive content?
    \answerYes{Yes. Please see Section~\ref{sec:ethics}.}
\item If you are curating or releasing new datasets, did you discuss how you intend to make your datasets FAIR?
\answerNA{N/A}
\item If you are curating or releasing new datasets, did you create a Datasheet for the Dataset? 
\answerNA{N/A}
\end{enumerate}

\item Additionally, if you used crowdsourcing or conducted research with human subjects, \textbf{without compromising anonymity}...
\begin{enumerate}
  \item Did you include the full text of instructions given to participants and screenshots?
    \answerNA{N/A}
  \item Did you describe any potential participant risks, with mentions of Institutional Review Board (IRB) approvals?
    \answerNA{N/A}
  \item Did you include the estimated hourly wage paid to participants and the total amount spent on participant compensation?
    \answerNA{N/A}
   \item Did you discuss how data is stored, shared, and deidentified?
   \answerNA{N/A}
\end{enumerate}

\end{enumerate}

\appendix
\section{Common Crawl Data Collection Policies}
\label{app:cc}
To the best of our knowledge, websites with specific ``nofollow'' or ``robots.txt'' rules are excluded from the public scrape of Common Crawl. Thus, the \ww dataset lacks links from specific proprietary websites like Facebook, Reddit, Twitter, Quora, YouTube, \etc, thereby limiting our sample of the Web to the surface web.

\section{Defining Webpage Segmentation Rules}
\label{app:rules}
While there exist many methods such as engineering shallow textual features \cite{kohlschutter2010boilerplate} and deep learning techniques \cite{vogels2018web2text} to perform webpage segmentation, here we rely on a simple and intuitive rule-based approach that we propose to segment a webpage into three classes, namely \emph{boilerplate}, \emph{main content}, and \emph{user generated}. This is primarily because of the following three requirements. First, owing to the sheer scale of the \ww dataset we need a webpage segmentation model that is efficient and scalable. Second, owing to the lack of readily available ground-truth annotations, the model should be able to learn to segment a webpage using a relatively small set of labeled data samples. Finally, the model should be effective---possessing high precision and recall---and generalizable to unseen data. We note that while the state-of-the-art machine learning approaches mentioned above are effective, they lack efficiency and scalability. On the contrary, the proposed rule-based approach stands strong on all three aforementioned requirements, and is therefore preferred in this work.

To actually define the structural rules, we follow an iterative process inspired by grounded theory, which involves a continuous cycle of sampling Wikipedia links, labeling them, learning the rules by examining the DOM tree, and refining the rules based on intermediate predictions. Four authors were involved in this process. Specifically, we do the following:
\begin{enumerate}
    \item Sample 1000 random Wikipedia links.
    \item Label each link as belonging to one of the three classes, namely boilerplate, main content, or user generated. 
    \item Learn the rules using the 1000 Wikipedia links (labeled in Step 2) and define a rule-based classification model.
    \item Sample 500 random Wikipedia links and apply the learned rules.
    \item Refine the rules based on false positives and negatives obtained on the 500 links (sampled in Step 4). 
    \item Repeat until convergence.
\end{enumerate}

Using the aforementioned iterative process, we learn structural rules that can be grouped into two broad HTML categories, namely tag-based and attribute-based. 
Then, for each Wikipedia link we iterate through its ancestors in the DOM tree to determine if a specific tag or attribute is included. 
Let us consider an example of using an attribute-based rule to classify a specific segment of a webpage as boilerplate. We found that ``blogroll'', which constitutes a set of links the author of a blog post tends to include on the sidebar of their page, is a strong determinant of the boilerplate segment within a webpage. 
To determine if ``blogroll'' is an ancestor of an embedded Wikipedia link, we iterate over the ancestors of the Wikipedia link in the DOM tree and conduct a free text search on the class and id attributes of its ancestor nodes for ``blogroll''. 
If found, the Wikipedia link is classified as being embedded in the boilerplate segment. 
A similar process is carried out for all other rules. 
While clashes in the predictions from rules occur rarely, whenever clashes do occur they are resolved as follows. 
We first check if the link can be classified to be embedded as a ``response''. If not, we next check the applicability of the ``boilerplate'' classification, failing which, the link is classified to be embedded within the ``main content'' segment of a webpage. 
Note that the final set of webpage segmentation rules learned using the aforementioned iterative process for the \ww dataset is presented in the README of our GitHub repository: \url{https://github.com/blind-anonymous/web2wiki}.

We evaluate the efficacy of the learned rules as follows. We sample 100 random Wikipedia links from each class using the predictions from the rule-based model, and label the sampled links to compute precision. Additionally, we sample 500 random Wikipedia links from the \ww dataset, apply the rules to obtain predictions, and label the sampled links to measure recall. Table~\ref{tab:order_prediction} presents the results.

\begin{table}[!t]
    \centering
    \begin{tabular}{l|r|r}
        \textbf{Segment} & \textbf{Precision} & \textbf{Recall} \\
        \hline
        Boilerplate & 0.65 & 0.78 \\
        Main content & 0.91 & 0.96 \\
        User generated & 0.96 & 0.57 
    \end{tabular}
    \caption{Evaluating the efficacy of the proposed rule-based model for determining the segment where a Wikipedia link is embedded within a webpage.}
    \label{tab:order_prediction}
\end{table}

\section{Defining a Taxonomy for Linking}
\label{app:taxonomy}
To define the taxonomy, the authors went through a series of iterative coding exercises to infer intents users may have in including a link to Wikipedia. 
In general, we found it difficult to infer intent without more direct approaches like surveys. This realization caused us to move away from trying to define a descriptive taxonomy, 
and instead understand the relationship between Wikipedia and the site that invokes it. Using this approach, we found two fundamental reasons for including a link: (1) delegate an explanation (content enrichment) and (2) attribution either through evidence or content. 

We then used Selenium to screenshot a sample of 500 Web links to Wikipedia and then went through a process of manually annotating them. 
Screenshots were used for annotation since the visual features were a better signal of intent than text alone. In total, 213 articles did not maintain the link or were functioning. For these cases, we relied on the archived HTML in our dataset and rendered the pages from the available code.

\section{Top domains}
Table \ref{tab:top_domains_full} summarizes the top 30 domains by number of links to Wikipedia.

\begin{table}[ht]
\centering
\begin{tabular}{|r|l|r|}
\hline
\textbf{Rank} & \textbf{Domain} & \textbf{\# Links} \\
\hline
\hline
1 & wordpress.com & 3149615 \\
2 & blogspot.com & 2588425 \\
3 & free.fr & 1205608 \\
4 & wikiwix.com & 1194518 \\
5 & atoogo.com & 1146676 \\
6 & appspot.com & 1051000 \\
7 & chinabuddhismencyclopedia.com & 1025668 \\
8 & wikitrans.net & 1016960 \\
9 & portalfield.com & 914446 \\
10 & wikigallery.org & 592547 \\
11 & fandom.com & 557148 \\
12 & dbpedia.org & 484611 \\
13 & iedit.nu & 408697 \\
14 & coinshome.net & 396156 \\
15 & pungenerator.org & 393664 \\
16 & goo.ne.jp & 365376 \\
17 & qaz.wiki & 346522 \\
18 & leparisien.fr & 316862 \\
19 & tsujitadao.jp & 312781 \\
20 & flintservice.org & 295047 \\
21 & pskreporter.de & 294224 \\
22 & theyworkforyou.com & 279800 \\
23 & ria.ru & 276521 \\
24 & workers.dev & 269546 \\
25 & austria-forum.org & 269009 \\
26 & wikirank.net & 258150 \\
27 & officeequipmenthub.us & 245752 \\
28 & syoboi.jp & 236344 \\
29 & bingj.com & 228153 \\
30 & the2010s.com & 225107 \\
\hline
\end{tabular}
\caption{Top 30 linked domains by number of incoming links---excluding \textit{wikipedia.org} and \textit{wikidata.org}.}
\label{tab:top_domains_full}
\end{table}

\end{document}